\documentclass[final,5p,times,twocolumn]{elsarticle}

\usepackage{amssymb}
\usepackage{booktabs}

\usepackage{lineno}
\usepackage{amsmath,amssymb}
\usepackage{amscd,float,array,bm}
\usepackage{color}

\journal{Physica D: Nonlinear Phenomena}

\newcommand{\Ree}{{\rm Re}}
\newcommand{\Ha}{{\rm Ha}}
\newcommand{\Rm}{{\rm Rm}}
\newcommand{\Pm}{{\rm Pm}}
\newcommand{\er}{{\bf\hat e_r}}

\newcommand{\ez}{{\bf\hat e_z}}
\newcommand{\ve}{{\mathbf{v}}}
\newcommand{\bv}{{\bf B}}

\newcommand{\rv}{\ensuremath{\mathbf{r}}}
\newcommand{\lp}{\ensuremath{\left(}}
\newcommand{\rp}{\ensuremath{\right)}}

\begin{document}

\begin{frontmatter}



\title{Long term time dependent frequency analysis of
  chaotic waves in the weakly magnetized spherical Couette system}

\author{Ferran Garcia}
\author{Martin Seilmayer}
\author{Andr\'e Giesecke}
\author{Frank Stefani}
\address{Helmholtz-Zentrum Dresden-Rossendorf, Bautzner
 Landstra\ss e 400, D-01328 Dresden, Germany}

\begin{abstract}

The long therm behavior of chaotic flows is investigated by means of
time dependent frequency analysis. The system under test consists of
an electrically conducting fluid, confined between two differentially
rotating spheres. The spherical setup is exposed to an axial magnetic
field. The classical Fourier Transform method provides a first
estimation of the time dependence of the frequencies associated to the
flow, as well as its volume-averaged properties. It is however unable
to detect strange attractors close to regular solutions in the
Feigenbaum as well as Newhouse-Ruelle-Takens bifurcation scenarios. It
is shown that Laskar's frequency algorithm is sufficiently accurate to
identify these strange attractors and thus is an efficient tool for
classification of chaotic flows in high dimensional dynamical
systems. Our analysis of several chaotic solutions, obtained at
different magnetic field strengths, reveals a strong robustness of the
main frequency of the flow. This frequency is associated to an
azimuthal drift and it is very close to the frequency of the
underlying unstable rotating wave. In contrast, the main frequency of
volume-averaged properties can vary almost one order of magnitude as
the magnetic forcing is decreased. We conclude that, at the moderate
differential rotation considered, unstable rotating waves provide a
good description of the variation of the main time scale of any flow
with respective variations in the magnetic field.
\end{abstract}



\begin{keyword}
Frequency Analysis \sep Chaos \sep Magnetohydrodynamics


\end{keyword}

\end{frontmatter}


\section{Introduction}

The magnetized spherical Couette (MSC) system -a liquid metal within
two differentially rotating spheres subject to a magnetic field-
represents one of the fundamental problems for studying
three-dimensional magnetohydrodynamic (MHD) instabilities
(\citet{HoSk01,Hol09,GJG11,Kap14,GaSt18}). The coupled effects of
rotation, magnetic fields and spherical geometry, are indeed common in
a wide range of processes occurring in celestial
objects (\citet{DoSo07,MoDo19}), including the generation of the
Sun's (\citet{Rud89}) and the Earth's magnetic fields (\citet{Jon11}), or
the transport mechanisms in accretion disks around black holes, stars,
and protoplanetary disks (\citet{JiBa13}). The latter have been
interpreted in terms of the magnetorotational instability
MRI (\citet{BaHa91}) which is nowadays considered the best explanation.

Starting with the work by \citet{BaHa91} the occurrence of the MRI has
been studied in great detail including numerical and experimental
work. Experimental investigations of the MRI were conducted at the
Helmholtz-Zentrum Dresden-Rossendorf (HZDR) using the GaInSn liquid
metal alloy within two rotating cylinders
(\citet{SGGRSSH06,SGGHPRS09,SGGGSGRSH14}), and in Maryland
(\citet{SMTHDHAL04}) with liquid sodium in spherical geometry. The
latter experiment by \citet{SMTHDHAL04} motivated the recent numerical
studies of \citet{Hol09} and \citet{GJG11} which, however, did not
interpret the observed instabilities as MRI but as typical
instabilities in magnetized spherical Couette (MSC) flows.

To shed light onto this controversy, the HEDGEHOG experiment
(Hydromagnetic Experiment with Differentially Gyrating sphEres HOlding
GaInSn) has been designed, at HZDR, to describe three-dimensional
magnetohydrodynamic instabilities, which are related to the
hydrodynamic jet instability, the return flow instability and the
Kelvin-Helmholtz-like Shercliff layer instability (see \citet{KKSS17}
and the references therein).  These instabilities have been studied in
the past (e.\,g. \citet{Hol09,GJG11,TEO11,Kap14}) by means of direct
numerical simulations (DNS) of the MSC system, and their
spatio-temporal symmetries and nonlinear dynamics have been recently
described in terms of bifurcation and dynamical systems theory by
\citet{GaSt18} and \citet{GSGS19,GSGS20}. We refer to the introductory
sections of these latter studies for a detailed summary and references
on the numerical studies in the field.

The MSC system is {\bf{SO}}$(2)\times${\bf{Z}}$_2$-equivariant,
i.\,e., invariant by azimuthal rotations and reflections with respect
to the equatorial plane, and thus a rich variety of nonlinear dynamics
is expected (\citet{CrKn91}) thanks to flow bifurcations occurring as
the parameters $\Ree$ (the Reynolds number measuring rotation rates)
and $\Ha$ (the Hartmann number measuring magnetic field strength) are
varied. Bifurcations occurring in systems with symmetry have been
largely studied in the past
(e.\,g. \citet{CrKn91,Ran82,GLM00,GoSt03}). In the particular case of
the MSC system at moderate $\Ree=10^{3}$ and $\Ha<80$, the numerical
continuation of rotating waves, the theoretical description of
modulated rotating waves and the appearance of complex waves and
chaotic flows have been recently presented in the studies of
\citet{GaSt18}, \citet{GSGS19}, and \citet{GSGS20}, respectively.  The
present study extends these previous works by analyzing the long term
behavior of the flows, with special focus on the estimation of the
main time scales involved in chaotic flows. The numerical approach
relies on a time dependent frequency spectrum analysis of very long
time series, including global as well as local flow
properties. Laskar's algorithm (\citet{Las90,LFC92,Las93}),
implemented in the SDDSToolkit (\citet{BESS17}), provides a useful
tool for an accurate determination of the fundamental frequencies of a
time series. Moreover, the study of the time dependent spectrum
provides an estimation of the diffusion of the orbit in the phase
space (\citet{Las93b}) and thus can be used to identify chaotic flow
behaviour and to study global dynamics. There exist other even more
accurate algorithms for the determination of fundamental frequencies,
for instance those based on collocation methods described in
\citet{GMS10,GMS10b} and the references therein, which have been used
successfully as dynamical indicators. The idea of the analysis of the
time-frequency dependence is common (see for instance the work by
\citet{DjRu08} or the very recent comparison in \citet{VNROBT20}) to
assess the chaotic behavior of a nonlinear system.

We recall that for a more complete description of chaos the
computation of the so called Lyapunov characteristic exponents LCE
(\citet{Ose68,BGGS80,GrLe91}) must be performed and that for this
purpose time series tools are available (e.\,g. \citet{HKS99}). The
latter are based on phase space reconstruction using the method of
delays (\citet{Tak81}) and require the adjustment of several
parameters such as the embedding dimension or the time delay.  A
comprehensive analysis of methods for computing LCE, including those
based on direct time integrations of the evolution equations and those
based on time series, has been recently performed in \citet{AKEDBK18}
for simple systems. In addition, the study of \citet{AKEDBK18}
provides a description of the dynamics in terms of Fourier spectra and
Gauss wavelets. In a subsequent study (\citet{AKEDBK18b}) the
comparison was extended to a system with $O(10^2)$ degrees of freedom
illustrating main problems and difficulties of LCE estimation from a
time series. In comparison with those techniques, Laskar's analysis
can be applied in a more straightforward manner and only requires to
control the accuracy of the obtained frequencies.

With the present study we demonstrate the applicability of Laskar's
algorithm, a highly accurate tool for the determination of fundamental
frequencies, for identifying chaotic motions in a dissipative
dynamical system with a large number $\sim O(10^5)$ of degrees of
freedom, due to the spatial discretization of partial differential
equations. This is demonstrated for the first time in the context of an
MHD problem in spherical geometry. In addition, the study is based on
very long time evolutions (more than one order of magnitude larger
than the previous studies in the MSC context) which is a challenging
task given the dimension of the problem. With the analysis of the time
dependent spectrum for two different routes to chaos, the main result
found is a strong robustness of the temporal scale associated to an
azimuthal flow drift, even for highly oscillatory chaotic flows.

The structure of the paper is as follows: In \S\ \ref{sec:mod} the
problem and the numerical method used to integrate the model equations
are formulated, and the data used for the spectral analysis is
described. In \S\ \ref{sec:freq} a study of the accuracy for the
frequency determination and the set-up for the time dependent spectra
is provided. The results are discussed in \S\ \ref{sec:res},
considering the Feigenbaum (\citet{Fei78}) as well as
Newhouse-Ruelle-Takens (\citet{NRT78}) routes to chaos, and finally in
\S\ \ref{sec:sum} the paper closes with a discussion on the main
results obtained.

\section{The model and methods}
\label{sec:mod}

\begin{figure}[h!]
\begin{center}
  \includegraphics[scale=0.9]{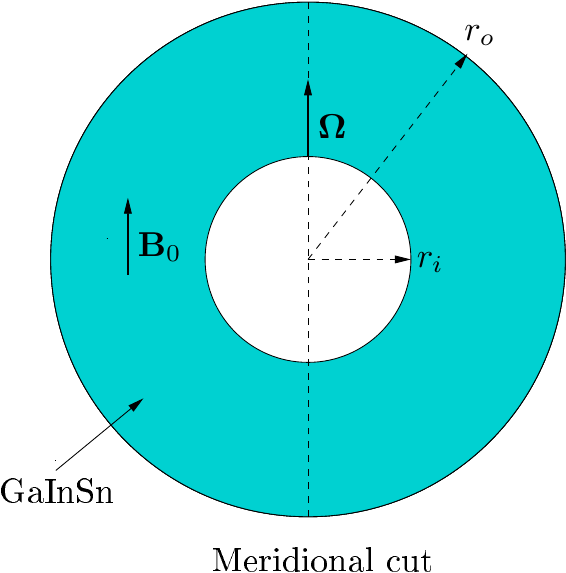}
\end{center}
\caption{Geometrical configuration of the magnetized spherical Couette (MSC) problem.}
\label{fig:geom}    
\end{figure}

In the HEDGEHOG experiment a liquid metal (GaInSn) fills the gap
between two spheres of radius $r_i$ and $r_o$ with
$\chi=r_i/r_o=0.5$. The inner sphere is rotating with constant
velocity $\Omega$ around the vertical axis $\ez$ while the outer is at
rest. In addition, an axial magnetic field of amplitude $B_0$ is
applied to the system, see Fig.~\ref{fig:geom}, and insulating
boundary conditions are considered for the magnetic field outside the
fluid region (e.\,g. \citet{HoSk01}).

The mathematical formulation of the problem relies on the
inductionless approximation of the Navier-Stokes and induction
equations (\citet{HoSk01}).  Considering $\eta$ as the magnetic
diffusivity, $\nu$ as the kinematic viscosity, and $d=r_o-r_i$, the
inductionless approximation remains valid when the magnetic Reynolds
number $\Rm=\Omega r_i d/\eta$ is small, $\Rm\ll 1$.  In case of the
HEDGEHOG experiment the GaInSn eutectic alloy (\citet{PSEGN14}) has
very low magnetic Prandtl number $\Pm=\nu/\eta\sim O(10^{-6})$ and the
values for the Reynolds numbers are moderate $\Ree=\Omega r_i
d/\nu\sim 10^3$. This means that $\Rm=\Pm\Ree \sim 10^{-3}$ and thus
the inductionless approximation is valid.

By scaling the length, time, velocity and magnetic field with
$d=r_{o}-r_{i}$, $d^2/\nu$, $r_{i}\Omega$ and $B_0$, respectively, the
equations of motion become
\begin{align}
  \partial_t\ve+\Ree\lp\ve\cdot\nabla\rp\ve &=
-\nabla p+\nabla^2\ve+\Ha^2(\nabla\times {\bf b})\times\ez, \label{eq:mom_less}   \\
 0& = \nabla\times(\ve\times\ez)+\nabla^2{\bf b}, \label{eq:ind_less}\\
\nabla\cdot\ve=0, &\quad \nabla\cdot{\bf b}=0.\label{eq:div}
\end{align}
where $p$ is the dimensionless pressure containing all the potential
forces, $\ve$ is the velocity field and ${\bf b}$ is the magnetic
field perturbation of the axially applied field $\bv=\ez+\Rm\,{\bf
  b}$. The no-slip ($v_r=v_\theta=v_\varphi=0$) and constant rotation
($v_r=v_\theta=0,~v_\varphi= \sin{\theta}$) conditions are imposed on
the boundary at $r=r_o$ and $r=r_i$, respectively.  For the magnetic
field the exterior regions are assumed to be insulating, as it is the
case for the HEDGEHOG experiment. The magnetic field boundary
conditions are formulated in terms of the spherical harmonics (see
\citet{HoSk01} for full details). The system is governed by 3
non-dimensional numbers:
\begin{equation*}
  \Ree=\frac{\Omega r_i
    d}{\nu} , \quad
\Ha=\frac{B_0 d}{\sqrt{\rho\nu\mu_0\eta}}, \quad\text{and}\quad
  \chi=\frac{r_i}{r_o},
\end{equation*}
with $\mu_0$ being the magnetic permeability for free-space and $\rho$
being the density of the fluid. The parameters selected for the
present study, $\chi=0.5$, $\Ree=10^{3}$ and $\Ha<6$, are in
accordance with the typical operating parameters of the HEDGEHOG
experiments, in which $\eta=0.35,0.5$, $\Ree\in[10^3,10^4]$ and
$\Ha<10^3$.

The pseudo-spectral method for the numerical solution of the governing
equations is briefly described in the following. For full details we
refer to \citet{GaSt18} and references therein. The divergence-free
velocity field
$\ve=\nabla\times\lp\Psi\rv\rp+\nabla\times\nabla\times\lp\Phi\rv\rp$
is expressed as a sum of the toroidal, $\Psi$, and poloidal, $\Phi$,
potentials, with $r = r~\er$ being the position vector. For the radial
coordinate a collocation method on a Gauss--Lobatto mesh of $N_r$
points is employed. For the angular coordinates the scalar potentials
are expanded in spherical harmonic series up to degree
$L_{\text{max}}$ and order $M_{\text{max}}=L_{\text{max}}$:
\begin{eqnarray}
 (\Psi,\Phi)(t,r,\theta,\varphi)=\sum_{l=0}^{L_{\text{max}}}\sum_{m=-l}^{l}{(\Psi,\Phi)_{l}^{m}(r,t)Y_l^{m}(\theta,\varphi)},
\label{eq:pot}  
\end{eqnarray}
with $\Psi_l^{-m}=(\Psi_l^{m})^*$ and $\Phi_l^{-m}=(\Phi_l^{m})^*$,
$(\cdot)^*$ meaning complex conjugation. By choosing
$\Psi_0^0=\Phi_0^0=0$ the two scalar potentials are uniquely
determined. We recall that $Y_l^{m}(\theta,\varphi)=P_l^m(\cos\theta)
\text{e}^{im\varphi}$ is the spherical harmonic function with $P_l^m$
being the normalized associated Legendre functions of degree $l$ and
order $m$. High order implicit-explicit backward differentiation
formulas IMEX--BDF (\citet{GNGS10}) are used for the time
integration. The nonlinear terms are considered explicitly to avoid
the solution of a nonlinear system at each time step. An explicit
treatment of the Lorenz force term facilitates the implementation of
the linear solver, but may lead to smaller time integration steps
($\Delta t$) in comparison with an implicit treatment. However, this
is not critical as moderate $\Ha$ are considered in the present study.

Two different diagnostics are considered for the analysis of the
DNS. First, the time series of the radial velocity $v_r$ picked up at
the point $(r,\theta,\varphi)=(r_i+0.5d,\pi/8,0)$, which is a local
measure that reflects the time scales of the flow. Second, the time
series of the volume-averaged kinetic energy $K$, defined as
\begin{equation}
K=\frac{1}{2{\mathcal V}}\int_{\mathcal V} \ve
\cdot \ve \;dv,
\label{eq:ener_dens}
\end{equation}
with $\mathcal V$ being the volume of the shell and $\ve$ being the
velocity field, provides a global measure. Instead of considering $K$
for the total flow, we compute the kinetic energy $K_m$ defined by
only employing the spherical harmonic amplitudes $\Psi^m_l$ and
$\Phi^m_l$ with order $m$ and degree $l$ satisfying $|m|\le l \le
L_{\text{max}}$. This provides an idea on the distribution of kinetic
energy among the different azimuthal modes $m$. For the flows we are
analyzing (see \citet{GSGS20}) there exists an $m_{\text{max}}$ with
$\overline{K}_{m_{\text{max}}}\gg \overline{K}_m$, $1\le m\le
L_{\text{max}},~\text{and}~m\ne m_{\text{max}}$, with the over-line
representing a time average. We note that then the flow will exhibit
$m_{\text{max}}$ vortices. If in addition the flow has $m_d$-fold
azimuthal symmetry, then it is unaffected by azimuthal rotations
multiples of $2\pi/m_d$ and the spherical harmonic amplitudes with
azimuthal wave numbers being multiples of $m_d$ are the only nonzero
ones in Eq.~(\ref{eq:pot}). Notice that if the azimuthal symmetry is
$m_d=1$ all the spherical harmonics amplitudes are considered.

\section{Frequency analysis}
\label{sec:freq}

For an accurate determination of the fundamental frequencies of a time
series, Laskar's method (\citet{Las93}) of numerical analysis of
fundamental frequencies (NAFF) is employed. This method, implemented
in the SDDSToolKit (\citet{BESS17}), involves a von-Hann-window and
FFT together with a numerical optimization of the difference between
the signal and exponential functions of time. Concretely, given a time
series $(t_i,p(t_i))$, $t_i=t_{i-1}+\Delta t$, $i=1,\ldots,N$ of a
quasiperiodic function $p$, defined on a time interval $[0,T]$,
Laskar's algorithm provides the decomposition of $p(t)$ on the basis
$\text{e}^{-\text{i}f_jt}$ which is nonorthogonal on a finite time
window, computing the frequencies $f_j$, $j=1,...,M$, with an
iterative algorithm. It first starts by finding the maximum term of
the FFT of the time series.  The corresponding frequency, $f$, is
refined to obtain $f_1$ by maximizing the power spectrum $$\int_0^T
p(t)\text{e}^{-\text{i}ft}H(t)dt,$$ where $H(t)=1+\cos(\pi t/T)$ is
the von-Hann-window filter (helps to reduce the coupling effect of
other frequencies). Once $f_1$ is found, the corresponding term is
removed from the time series and the process is repeated to find
$f_2$. After finding each $f_j$ we note that since the basis functions
$\text{e}^{-\text{i}f_jt}$ are not orthogonal an intermediate step of
orthogonalization (Gauss algorithm) is required to compute the
amplitudes. The algorithm stops whenever the new frequency $f_k$
satisfies $|f_k-f_j|< 1.5\times 2\pi/T$, for any $j<k$, which
corresponds to the band limitation of an FFT with a von-Hann-window
filter.

\subsection{Accuracy estimation}
\label{sec:ac_freq}

In order to estimate the accuracy of Laskar's algorithm for the
determination of the frequency with the largest amplitude, a rotating
(also travelling) wave (RW) with azimuthal symmetry $m=4$ and rotating
frequency $\omega$ is considered (i.\,e. a periodic flow whose temporal
dependence can be described as $u(t,r,\theta,\varphi)\equiv
u(r,\theta,\varphi-\omega t)$). The parameters of this RW are
$\chi=0.5$, $\Ree=10^3$ and $\Ha=3.7766571$.

Because a RW is a periodic orbit, it can be obtained by means of a
continuation method (\citet{GaSt18}) and its rotating frequency
estimated up to a prescribed tolerance. Specifically, we solve a
nonlinear system which determines a single RW defined by $(u,\tau,p)$,
with $\tau=2\pi/(m\omega)$ being the period, at a parameter
$p=\Ha$. The system is
\begin{equation}
H(u,\tau,p)= \left(
\begin{array}{c}
u-\phi(\tau,u,p)\\
g(u)\\
m(u,\tau,p)\\
\end{array}
\right)
=0,
\label{eq:H_eq}
\end{equation}
where $\phi(\tau,u,p)$ is a solution of
Eqs.~(\ref{eq:mom_less})-(\ref{eq:div}) at time $\tau$ and initial
condition $u$ for fixed $p$. The condition $g(u)=0$ is selected to fix
the undetermined phase of the RW and $m(u,\tau,p)$ is the
pseudo-arclength condition of the continuation method. This system is
solved by employing a Newton-Krylov procedure with tolerance
$10^{-8}$. The method is matrix-free and so does not require the
explicit computation of the Jacobian $D_{(u,\tau,p)}H(u,\tau,p)$, but
only its action on a given vector (see \citet{GaSt18} for full
details).

Once the RW with $m=4$ at $\Ha=3.7766571$ has been obtained with the
Newton-Krylov procedure, a direct time integration of the MSC
equations is performed to obtain the time series of the radial
velocity $v_r$ picked up at the point
$(r,\theta,\varphi)=(r_i+0.5d,\pi/8,0)$ for which Laskar's algorithm
is applied subsequently. This particular location allows to measure
the meridional circulation of the flow at high latitudes. Because the
radial velocity amplitude is significant on a wide region in the bulk
of the shell (see e.\,g. Fig. 12 in \citet{GSGS20}) the results should
not depend on the measurement position. We have checked this by also
considering the point
$(r,\theta,\varphi)=(r_i+0.85d,3\pi/8,\pi/2)$. The step for the time
integration is the same, $\Delta t=5\times 10^{-6}$, as that used for
the time integration within Newton's method. We note that this $\Delta
t$ provides enough accuracy (we have used a 4th order time integration
scheme) with errors less than $10^{-8}$ because otherwise Newton's
method does not converge. The use of high order time integration
methods is recommended when a highly accurate time integration is
required (\citet{GNGS10}).

Newton's method (with tolerance $10^{-8}$) gives $\omega=138.09097$
which corresponds to the frequency $f=m\omega/2\pi=87.91144$ whereas
Laskar's algorithm result is $f=87.91145$ for $T\ge0.3$, $T$ being the
time interval of the time series from which the frequency $f$ has been
computed. The sampling time interval is $\Delta
t_{\text{samp}}=10^{-4}$ dimensionless time units. The case yields
$f=87.91068$ for $T=0.1$, but $f=87.91151$ for $T=0.2$. By decreasing
$\Delta t_{\text{samp}}=10^{-5}$ the same results are obtained, but
for $\Delta t_{\text{samp}}=2\times 10^{-4}$ the accuracy is degraded
to $f=87.91167$ even for $T=20$. The accuracy provided by Laskar's
algorithm is estimated to be $O(1/T^3)$ (e.\,g. \citet{Las93b})
contrasting the $O(1/T)$ estimation for the classical FFT.  In the
case of computing $f$ with FFT, with $\Delta t_{\text{samp}}=10^{-4}$,
we obtain $f=87.9000$ for $T=20$ and $f=87.91666$ for $T=60$, meaning
that a time window $T\ge 60$, which is large for a simple periodic
time series, has to be considered in the FFT to detect changes in the
frequency which are below $1$\%. As it will be evidenced latter, this
makes the FFT unpractical to detect the chaotic flows studied here.

\subsection{Time dependent frequency spectrum}
\label{sec:ti_de_freq}

Our analysis is based on very long time integrations up to a
  final time $T_f=100$ (in dimensionless time units), which
corresponds to $2\times 10^7$ time integration steps ($\Delta
t=5\times 10^{-6}$). This is a challenging task because of the large
dimension $n\sim O(10^5)$ of the ODE system due to the spatial
discretization of the MSC governing equations. 

Given a flow initial condition $u\in\mathbb R^n$, which will be
detailed in the next section, the frequency $f$ of maximum amplitude
is computed from a time window $[t,t+T]\subset[0,T_f]$ of the time
series. According to \citet{Las93b} this provides the map
\begin{eqnarray*}
F_T:  \mathbb R^n\times \mathbb R&\hspace{-0.8cm}\xrightarrow{\hspace{1.cm}} \mathbb R\\
  (u,t) &\xrightarrow{\hspace{1.cm}} f(u,t)
\end{eqnarray*}
which can be used for the analysis of the diffusion of the orbit
$u\in\mathbb R^n$ with respect to time, and thus infer the regular or
chaotic behavior of the orbit. If the flow is quasiperiodic $F_T$ is
a constant function of time whereas for chaotic flows $F_T$ varies
indicating the diffusion of the orbit in phase space.

As noticed in \citet{Las93b} the frequencies are computed up to a
certain accuracy, $\epsilon_f$, depending on the solution $u$ and the
time window $T$. For the case of a rotating wave, we have shown in the
previous section that this accuracy for Laskar's algorithm is about
$\epsilon_f=10^{-5}$, even for very small time windows $T=0.3$. For
the analysis of quasiperiodic and chaotic flows we may assume slightly
larger discrepancies $\epsilon_f$. Then, flows are considered to be
regular if $F_T$ is constant within $\epsilon_f$ of accuracy,
otherwise to be chaotic. Following \citet{Las93b} we also estimate the
instantaneous diffusion rate as $\delta
F_T(u,t)=|F_T(u,t+T)-F_T(u,t)|$. The diffusion of the orbit is
nonzero, i.\,e. the flow is nonregular, when $\delta
F_T(u,t)>\epsilon_f$. For the analysis of the flows presented in this
study we evaluate $F_T(u,t)$ at the time instants $t_i=0.1(i-1)\leq
T_f-T$ and $\delta F_T(u,t)$ at the same time instants but for
$t_i\leq T_f-2T$.

For the analysis, either with Laskar's or FFT algorithm, of all the
time series considered in this study the setup is the following: A
sampling time step $\Delta t_{\text{samp}}=10^{-4}$ has been used as
it provides the best accuracy from the estimation given in
Sec. \ref{sec:ac_freq}. In addition, several sizes of the time window
$T=1,~2.5,~5,~10,~20$ and $40$ have been considered to check that the
results do not depend on the particular choice of the time window
size.

\section{Results: Analysis of chaotic flows}
\label{sec:res}

Several flow realizations that correspond to two different routes to
chaos (\citet{Eck81}) are studied in this section. The first scenario
is in accordance with the Feigenbaum route (\citet{Fei78}) in which
chaotic flows are developed after a sequence of period doubling
bifurcations.  The second scenario corresponds to the
Newhouse-Ruelle-Takens route (\citet{NRT78}) in which strange
attractors develop from a sequence of Hopf bifurcations giving rise to
quasiperiodic flows. These two routes have been already identified and
described in the recent study of \citet{GSGS20} devoted to the MSC
problem in a parameter regime corresponding to the radial jet
instability, in which the magnetic effects are weak. The specific
problem parameters are $\chi=0.5$, $\Ree=10^3$ and $\Ha<4$.  The
description of \citet{GSGS20}, based on the study of bifurcation
diagrams and Poincar\'e sections, evidenced both
scenarios. Concretely, for the Feigenbaum scenario, the distance
between the successive period bifurcation points was used in
\citet{GSGS20} to estimate the Feigenbaum constant within $5$\% of
accuracy. In the case of the Newhouse-Ruelle-Takens scenario, the
different bifurcations giving rise to two and three-frequency flows,
before the appearance of chaos, were identified in \citet{GSGS20}.

For the present study we are interested in the description of long
term behavior of solutions belonging to these two scenarios which give
rise to chaotic flows.  We select three different flows at each of the
routes (Newhouse-Ruelle-Takens and Feigenbaum) presented in
\citet{GSGS20} and perform the analysis summarized in
Sec.~\ref{sec:ti_de_freq} to estimate the diffusion of the
corresponding orbits and the time variation of the main frequency $f$
(that with maximum amplitude) either of the time series of the radial
velocity or the time series of the $m=2$ volume-averaged kinetic
energy (already defined in Sec.\ref{sec:mod}). These time series are
obtained from direct numerical simulations as described in
\citet{GSGS20}. The spatial resolution requirements, $N_r=40$ and
$L_{\max}=84$, for solving the MSC equations have been already
validated in \citet{GaSt18} and, as commented in
Sec.~\ref{sec:ac_freq}, high order time integration is employed to
obtain accurate time series.

\subsection{Feigenbaum scenario}

\begin{figure}[b!]
\includegraphics[scale=0.95]{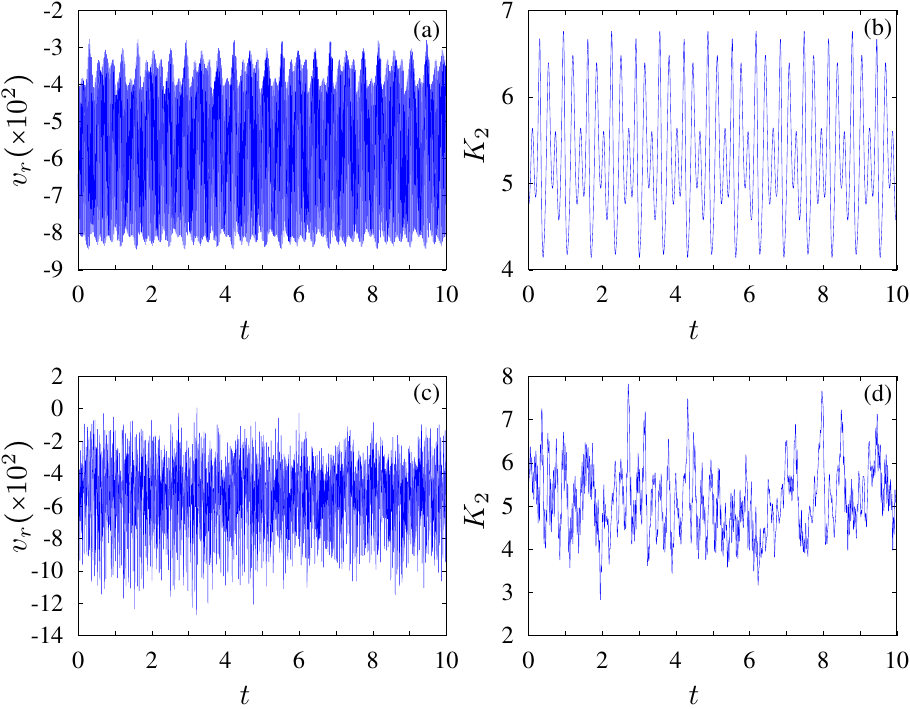}
\caption{(a,c) The time series of the radial velocity picked up at the
  point $(r,\theta,\varphi)=(r_i+0.5d,\pi/8,0)$. (b,d) The time series
  of the volume averaged kinetic energy of the $m=2$ component of the
  flow. (a,b) are for a regular flow with two fundamental frequencies
  at $\Ha=3.425$, and (c,d) are for a chaotic flow at $\Ha=0.7$.}
\label{fig:ti_se}    
\end{figure}

This section focuses on chaotic flows originating from a
period-doubling cascade of quasiperiodic flows with two fundamental
frequencies (i.\,e. two-tori) and with $m=2$ azimuthal symmetry. A
detailed analysis of this scenario was performed in \citet{GSGS20}. In
this latter study the Feigenbaum iterates
$\delta_i=(\Ha_{i+1}-\Ha_{i})/(\Ha_{i+2}-\Ha_{i+1})$, with
$\Ha_1=3.491$, $\Ha_2=3.423$, $\Ha_3=3.4073$, and $\Ha_4=3.4039$ being
the successive period-doubling bifurcation points, have been estimated
to be $\delta_1=4.33$ and $\delta_2=4.62$, in reasonable agreement
with the Feigenbaum constant $\delta=4.6692$.

For the analysis of the time dependence of frequency spectra in the
Feigenbaum route a quasiperiodic flow with two fundamental frequencies
at $\Ha=3.425$ and two chaotic flows at $\Ha=3.4$ and $\Ha=0.7$ are
selected. Notice that for $\Ha=3.425$ a period-doubling bifurcation
has already occurred and that the chaotic flow at $\Ha=3.4$ is close
to the onset of chaos (the last period doubling found is at
$\Ha_4=3.4039$). We have selected this regular solution to compare the
results with the two other chaotic flows along the Feigenbaum
route. The regular and chaotic flows at $\Ha=3.425$ and $\Ha=3.4$ have
$m=2$ azimuthal symmetry whereas the azimuthal symmetry of the chaotic
flow at $\Ha=0.7$ is $m=1$. As studied in \citet{GSGS20}, the chaotic
nature of the flows remains by decreasing $\Ha$ from $\Ha=3.4$. The
flows close to $\Ha=0$ are strongly oscillatory with $m_{\max}=2$.

\begin{figure}[h!]
\hspace{0.cm}\includegraphics[scale=1.04]{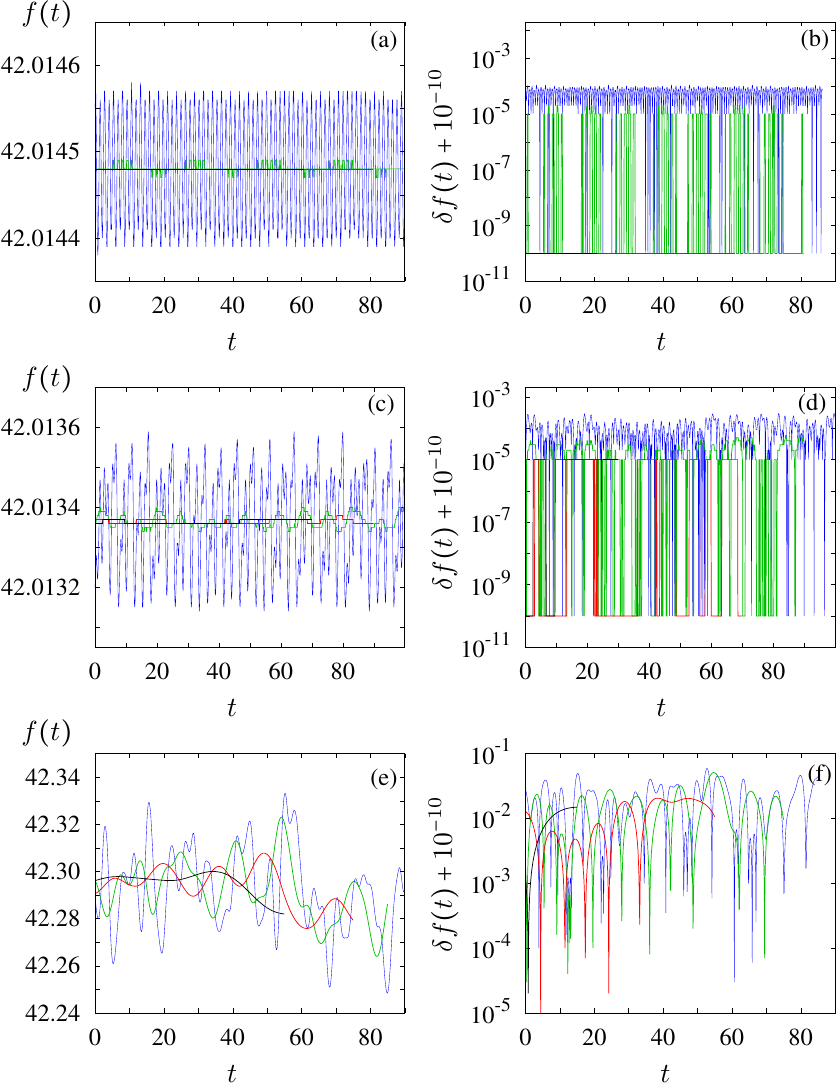}
\caption{Time dependent frequency spectrum based on Laskar algorithm
  (SDDSToolKit). The time series correspond to the radial velocity
  picked up at the point $(r,\theta,\varphi)=(r_i+0.5d,\pi/8,0)$.
  (a,c) Frequency with maximum amplitude versus time. (b,d) Time
  difference $\delta f(t)=|f(t+T)-f(t)|$ versus time
  (logscale). Different colors denote different lengths of the time
  series (blue $T=5$, green $T=10$, red $T=20$ and black
  $T=40$). Panels (a,b) are for a regular solution at $\Ha=3.425$,
  panels (c,d) are for a chaotic solution at $\Ha=3.4$, and panels
  (e,f) for a chaotic solution at $\Ha=0.7$.}
\label{fig:Lask_driftmov_PDC}   
\end{figure}

Figure~\ref{fig:ti_se} displays the time series of the radial velocity
$v_r$ and the volume averaged kinetic energy $K_2$ of the $m=2$
component of the flow (see Sec.~\ref{sec:mod}) for the regular and
chaotic flows at $\Ha=3.425$ and $\Ha=0.7$, respectively, in case of
the Feigenbaum scenario. For the 2-frequency solution (panels (a) and
(b)), the time series of $v_r$ exhibits a quasiperiodic behavior
whereas the time series of $K_2$ remains periodic, showing the
period-doublings. This is because the solution is a modulated rotating
wave (\citet{Ran82,GSGS19}) and one of the frequencies is associated
with the rigid rotation (azimuthal drift) of the flow patterns. By
azimuthally averaging the flow, only the frequency of modulation is
observed. The time series of the chaotic flow at $\Ha=0.7$ (panels (c)
and (d)) exhibit a clear chaotic behavior, but the small temporal
scale (associated to the azimuthal drift shown in panel (a)) of the
radial velocity still prevails.

\begin{figure}[h!]
\hspace{0.cm}\includegraphics[scale=1.01]{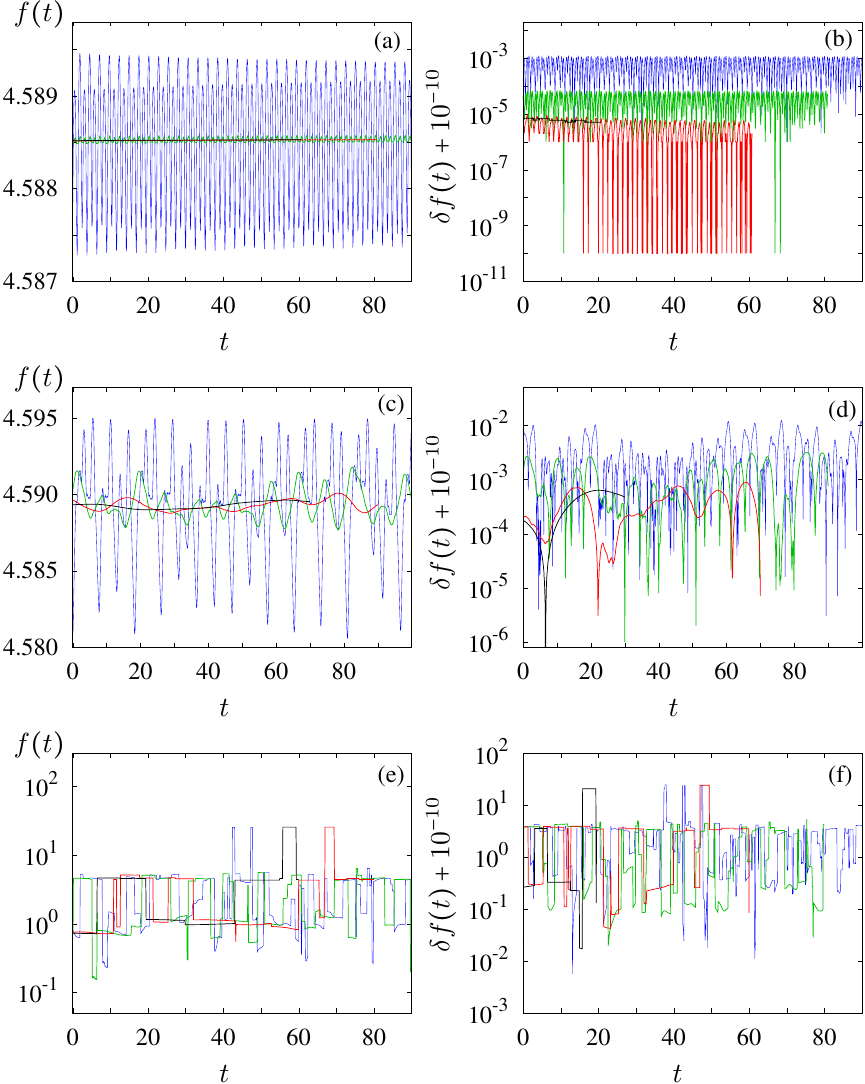}
\caption{Time dependent frequency spectrum based on Laskar algorithm
  (SDDSToolKit). The time series corresponds to the volume averaged
  kinetic energy of the $m=2$ component of the flow.  (a,c) Frequency
  with maximum amplitude versus time. (b,d) Time difference $\delta
  f(t)=|f(t+T)-f(t)|$ versus time (logscale). Different colors denote
  different lengths of the time series (blue $T=5$, green $T=10$, red
  $T=20$ and black $T=40$). Panels (a,b) are for a regular solution at
  $\Ha=3.425$, panels (c,d) are for a chaotic solution at $\Ha=3.4$,
  and panels (e,f) for a chaotic solution at $\Ha=0.7$.}
\label{fig:Lask_mov_PDC}   
\end{figure}

\begin{figure}[h!]
\begin{center}
\hspace{0.cm}  \includegraphics[scale=0.83]{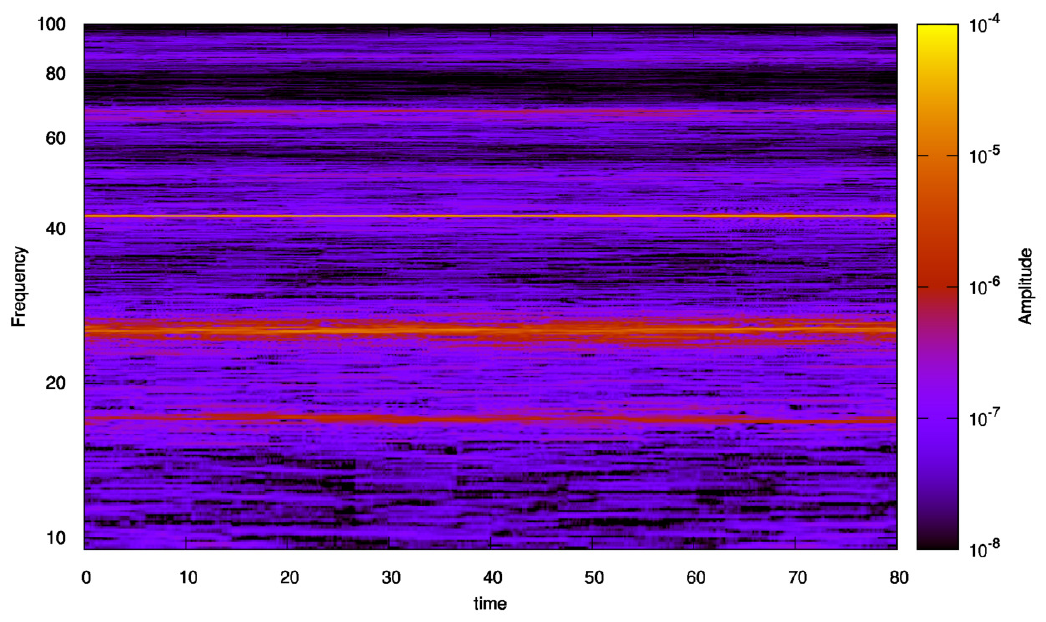}\\[0.mm]
\hspace{0.cm}  \includegraphics[scale=0.83]{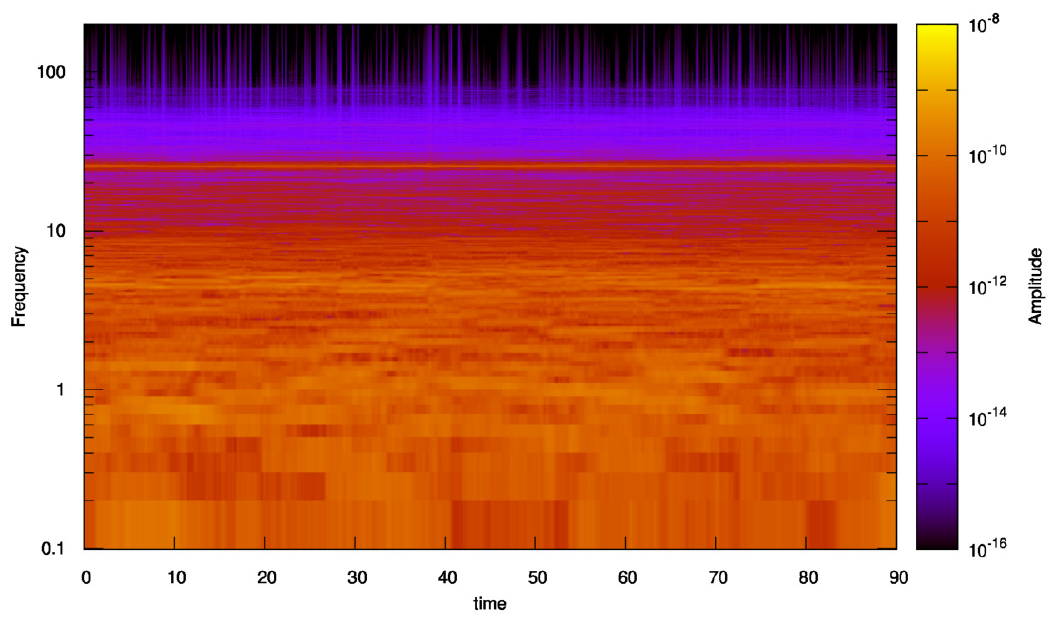}
\end{center}  
\caption{Time dependent frequency spectrum based on FFT on a time
  window $T=10$ for the chaotic flow at $\Ha=0.7$. The time series
  correspond to (a) the radial velocity picked up at the point
  $(r,\theta,\varphi)=(r_i+0.5d,\pi/8,0)$ and (b) the volume averaged
  kinetic energy of the $m=2$ component of the flow.  For the
  volume-averaged kinetic energy, the frequency with maximum amplitude
  varies on a broad range $f\in(0.1,100)$. In contrast, for the time
  series corresponding to the local measurement (radial velocity) the
  frequency with maximum amplitude remains constant ($f=42.3$) in the
  whole time range.}
\label{fig:fft_mov_Ch}   
\end{figure}

Figure~\ref{fig:Lask_driftmov_PDC}(a,b) illustrates the analysis of
the radial velocity time series of the regular solution at
$\Ha=3.425$, corresponding to a quasiperiodic flow with two
fundamental frequencies (a modulated rotating wave, see
\citet{GSGS19}) and with azimuthal symmetry
$m=2$. Figure~\ref{fig:Lask_driftmov_PDC}(a) provides $f(t)$ computed
using a time window of $T=5,10,20,40$ (the higher the amplitude of the
oscillations the smaller the time window) for the regular wave. As $f$
is computed from the time series of $v_r$ it corresponds to the
frequency of the azimuthal drift of the wave. A very weak time
dependence is observed which damps out by increasing $T$. For $T\ge
10$ the relative oscillations of $f$ are less than $10^{-5}$ and the
time difference $\delta f(t)=|f(t+T)-f(t)|\lesssim 10^{-4}$ (see
Fig.~\ref{fig:Lask_driftmov_PDC}(b)). We assume this values to be
valid for classifying this flow as regular, considering an accuracy
$\epsilon_f=10^{-4}$ for the frequency determination. We note that
although a value of $\epsilon_f=10^{-5}$ was achieved in
Sec.~\ref{sec:ac_freq} in the case of a rotating wave (i.\,e. a
periodic orbit), the regular solution now has 2 fundamental
frequencies which may increase the uncertainty in frequency
determination. We notice, however, that for $T\ge 20$ the value of $f$
is constant within $\epsilon_f=10^{-7}$.

Because the flow at $\Ha=3.4$ is close to the origin of
period-doubling chaos, the range of variation of $f$ and $\delta f$ is
small but relevant, providing a chaotic signature (see
Fig.~\ref{fig:Lask_driftmov_PDC}(c,d)). For the largest time window
considered, $T=40$, the difference value is $\delta t\le 10^{-5}$,
clearly larger than $\epsilon_f=10^{-7}$. The values for $f$ and
$\delta f$ for the chaotic flow at $\Ha=0.7$ (shown in
Fig.~\ref{fig:Lask_driftmov_PDC}(e,f)) are more pronounced but still
remain small. For instance, $f$ oscillates around its mean value with
less than $1\%$ for all considered time windows $T$, which indicates
the robust character of the frequency associated to the azimuthal
drift, even for this highly oscillatory flow.

The analysis for the volume averaged kinetic energy of the $m=2$
component of the flow, summarized in Fig.~\ref{fig:Lask_mov_PDC},
provides a better measure of chaotic behavior as the value of $\delta
f$ and the interval of variation of $f$ increase by one order of
magnitude with respect to the analysis of the radial velocity. As a
consequence different diffusion rates of the orbit emerge within the
phase space so that volume-averaging provides a better description of
these chaotic flows. We note that for the regular solution a slightly
noticeable transient can be identified on
Fig.~\ref{fig:Lask_mov_PDC}(a,b), because of the regular solution at
$Ha=3.425$ being close to the second period doubling bifurcation at
$\Ha_2=3.423$, so that long transients can be expected. Nevertheless
values of $\delta f<10^{-5}$, for $T\ge 20$, are obtained, which
supports our assumption of $\epsilon_f=10^{-5}$ for the largest time
windows. The value of $\delta f$ is clearly larger than this threshold
(for $T\ge 20$) when analyzing the chaotic solution at $\Ha=3.4$ (see
Fig.~\ref{fig:Lask_mov_PDC}(d)).

In contrast to the previous chaotic flows, the description for the
highly oscillatory flow at $\Ha=0.7$ is substantially different (see
Fig.~\ref{fig:Lask_mov_PDC}(e,f)). In this case, the frequency $f$
spans around two orders of magnitude and the values of $\delta f$
raise up to $O(10)$, accounting for a wide range of temporal scales in
volume-averaged quantities. We recall that this was not the case for
the main frequency of the radial velocity displayed in
Fig.~\ref{fig:Lask_driftmov_PDC}(e,f). To highlight the differences
between the main time scales of the flow and those of volume-averaged
quantities the time dependent spectrum based on FFT is provided in
Fig.~\ref{fig:fft_mov_Ch}. With the FFT analysis and a time window of
$T=10$ the frequency of maximum amplitude of the flow remains constant
at $f=42.3$ whereas a broad band of main frequencies is obtained in
case of the volume-averaged kinetic energy of the $m=2$ component of
the flow.

We note that the FFT analysis for the chaotic solution at $\Ha=3.4$
provides constant frequencies $f=42$ and $f=4.6$ for $v_r$ and $K_2$,
respectively. We recall that, as for Laskar's analysis, the time
windows $T=5,10,20$, and $40$, and sampling time step $\Delta
t_{\text{samp}}=10^{-4}$ are considered. Then, FFT analysis is unable
to detect the chaotic nature for solutions with $\Ha$ being close to
the onset of chaos. This is because for these flows $\delta f$ (see
either Fig.~\ref{fig:Lask_driftmov_PDC}(c,d) or
Fig.~\ref{fig:Lask_mov_PDC}(c,d)) is smaller than the accuracy
$O(1/T)>0.02$ achieved with the FFT.  This will be further evidenced
on the next section dedicated to the Newhouse-Ruelle-Takens scenario.

\subsection{Newhouse-Ruelle-Takens scenario}

\begin{figure}[b!]
  \hspace{0.cm}\includegraphics[scale=1.03]{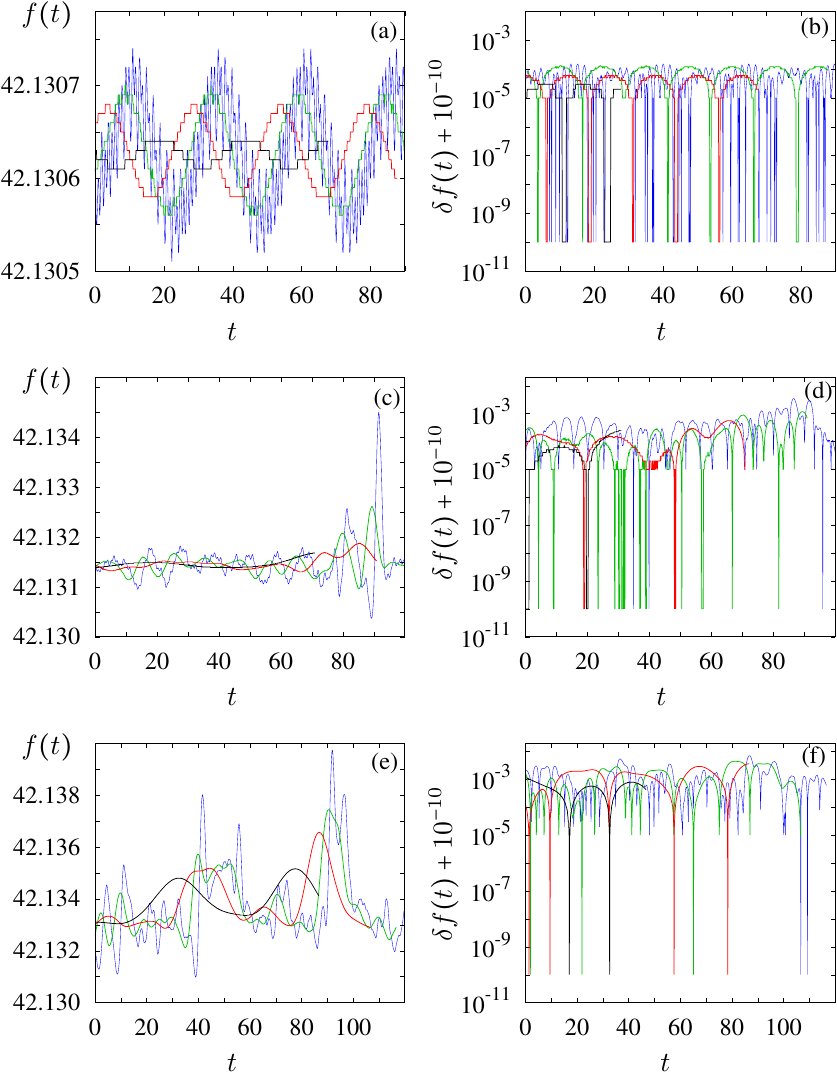}
\caption{Time dependent frequency spectrum based on Laskar's algorithm
  (SDDSToolKit). The time series correspond to the radial velocity
  picked up at the point
  $(r,\theta,\varphi)=(r_i+0.5d,\pi/8,0)$. (a,c,e) Frequency with
  maximum amplitude versus time. (b,d,f) Time difference $\delta
  f(t)=|f(t+T)-f(t)|$ versus time (logscale). Different colors denote
  different length of the time series (blue $T=5$, green $T=10$, red
  $T=20$ and black $T=40$). Panels (a,b) are for a regular solution at
  $\Ha=0.7$, panels (c,d) are for a chaotic solution at $\Ha=0.67$,
  and panels (e,f) are for a chaotic solution at $\Ha=0.63$.}
\label{fig:Lask_driftmov_RT}   
\end{figure}

One quasiperiodic flow with three fundamental frequencies at
$\Ha=0.7$, and two chaotic flows at $\Ha=0.67$ and $\Ha=0.63$ are
considered for the Newhouse-Ruelle-Takens scenario. They belong to the
same branch as described in \citet{GSGS20} with $m=1$ azimuthal
symmetry and $m_{\max}=2$. Bifurcation diagrams of two and three
frequency solutions and eventually chaotic flows, characteristic of
the Newhouse-Ruelle-Takens scenario, were analyzed in \citet{GSGS20}
for $\Ha<1$ in terms of Poincar\'e sections. We show in this section
that the appearance of chaos can be also evidenced by investigating
the time dependence of the main frequency obtained with Laskar's
procedure.

\begin{figure}[h!]
  \hspace{0.cm}\includegraphics[scale=1.03]{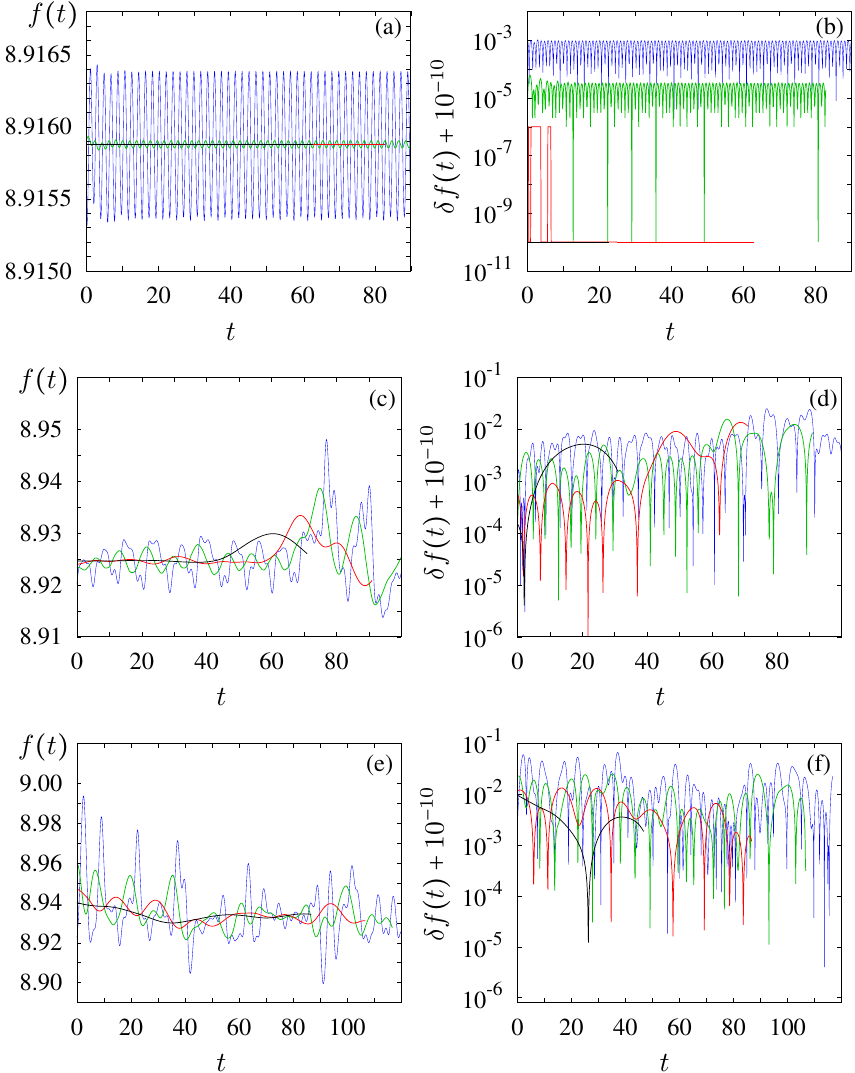}
\caption{Time dependent frequency spectrum based on Laskar's algorithm
  (SDDSToolKit). The time series correspond to the volume averaged
  kinetic energy of the $m=2$ component of the flow.  (a,c,e)
  Frequency $f$ with maximum amplitude versus time. (b,d,f) Time
  difference $\delta f(t)=|f(t+T)-f(t)|$ versus time
  (logscale). Different colors denote different length of the time
  series (blue $T=5$, green $T=10$, red $T=20$ and black
  $T=40$). Panels (a,b) are for a regular solution at $\Ha=0.7$,
  panels (c,d) are for a chaotic solution at $\Ha=0.67$, and panels
  (e,f) are for a chaotic solution at $\Ha=0.63$.}
\label{fig:Lask_mov_RT}   
\end{figure}

Figure~\ref{fig:Lask_driftmov_RT}(a) displays $f(t)$ for the regular
wave with 3 fundamental frequencies at $\Ha=0.7$. In comparison with
the regular solution of the Feigenbaum scenario, the variation of the
frequency with maximum amplitude $f$ and the value of $\delta f$ is
clearly larger for a fixed time window $T$.  The regular solution for
the Feigenbaum scenario has two fundamental frequencies whereas that
of the Newhouse-Ruelle-Takens scenario has three. This may be the
reason for the smaller value of $\epsilon_f$ achieved for the regular
solution in case of the Feigenbaum route.

For the chaotic flows at $\Ha=0.67$
(Figs.~\ref{fig:Lask_driftmov_RT}(c,d)) and $\Ha=0.63$
(Figs.~\ref{fig:Lask_driftmov_RT}(e,f)) the variation of $f$ and the
value of $\delta f$ is significant and the amplitude of their
oscillations is growing with time. For the chaotic flows the value of
$\delta f$ is at least one order of magnitude larger than for the
regular flow, although it is still small, indicating slow diffusion of
the orbit in the phase space. Indeed, as for chaotic flows in the
Feigenbaum scenario, the range of variation of $f$ is narrow which
indicates a nearly uniform azimuthal drift for these chaotic flows.

As in the Feigenbaum case, the frequency description considering the
$m=2$ volume-averaged kinetic energy, which is summarized in
Fig.~\ref{fig:Lask_mov_RT}, is even more clear than that corresponding
to the radial velocity. For the chaotic flows at $\Ha=0.67$ and
$\Ha=0.63$ the maximum value of $\delta f$ is larger than $10^{-2}$,
but the corresponding maximum value is smaller than $10^{-5}$ for the
regular solution at $\Ha=0.7$.  In addition, and in agreement with the
Feigenbaum scenario, for the chaotic flows at $\Ha=0.67$ and
$\Ha=0.63$ the value of $\delta f$ is significantly larger when
considering a volume-averaged measure than when considering a measure
of the flow itself. As discussed in the previous section this
indicates two very different diffusion rates of the orbit in the phase
space.

\begin{figure}[t!]
\begin{center}
\hspace{0.cm}\includegraphics[scale=0.83]{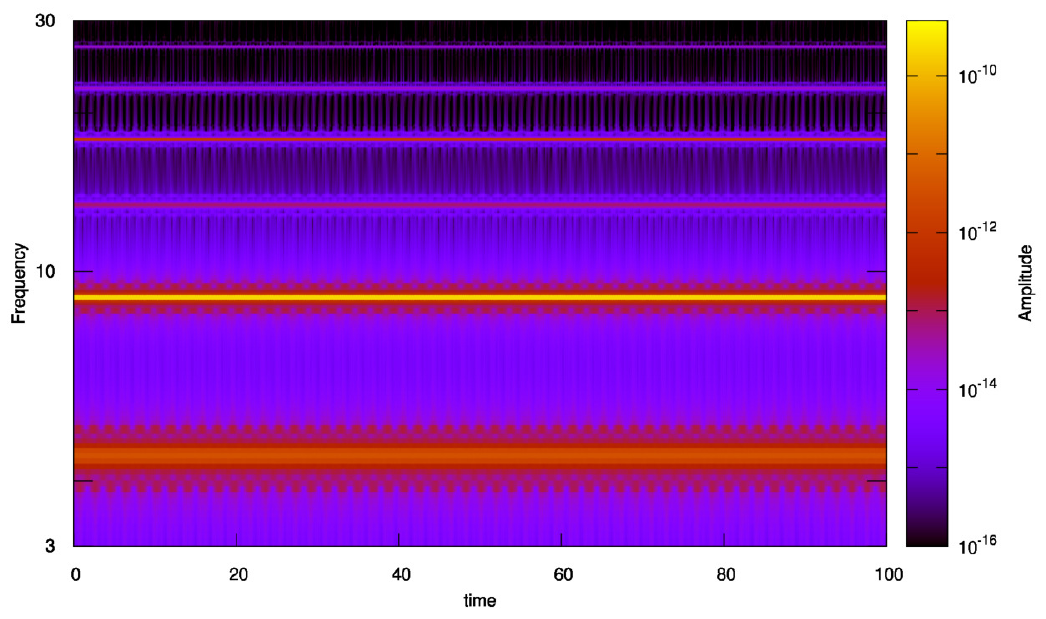}\\[0.mm]
\hspace{0.cm}\includegraphics[scale=0.83]{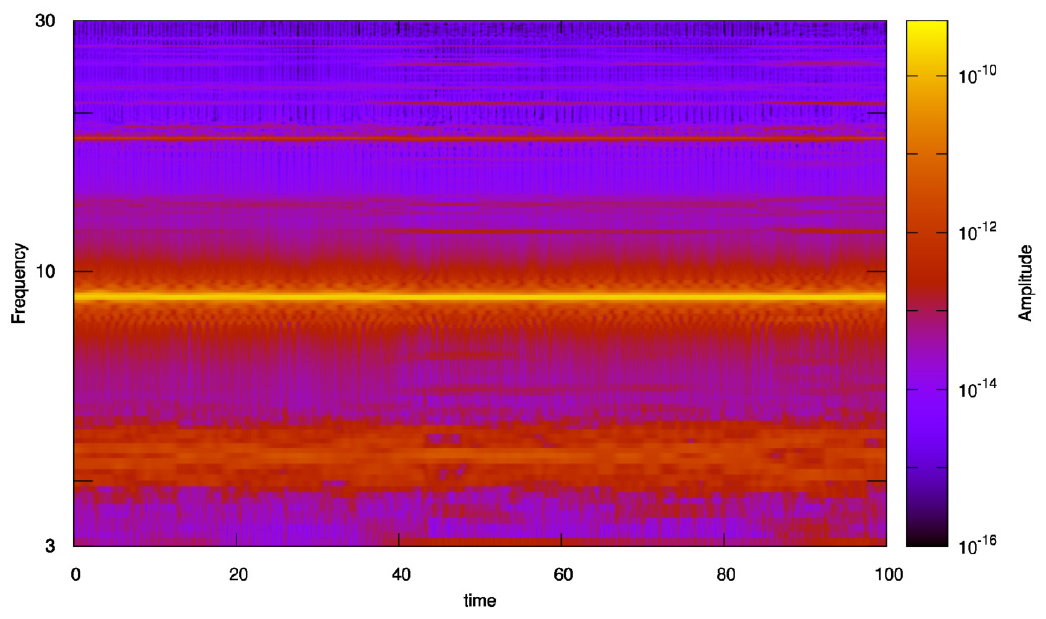}
\end{center}  
\caption{Time dependent frequency spectrum based on FFT on a time
  window $T=10$. The time series correspond to the volume averaged
  kinetic energy of the $m=2$ component of the flow. (a) Three
  frequency quasiperiodic flow at $\Ha=0.74$ and (b) Chaotic flow at
  $\Ha=0.63$. The frequency with maximum amplitude remains constant in
  the whole time range and is $f=8.9$, for both $\Ha=0.74$ and
  $\Ha=0.63$, and thus it does neither reveal the chaotic behavior for
  $\Ha=0.63$ nor it reflects the dependence of $f$ on $\Ha$.}
\label{fig:fft_mov_RT}   
\end{figure}

To highlight the superiority of Laskar's algorithm with respect to the
common FFT the moving FFT frequency spectrum of the volume-averaged
kinetic energy of the $m=2$ component of the flow is presented in
Fig.~\ref{fig:fft_mov_RT}(a,b). The top plot corresponds to a regular
flow at $\Ha=0.74$, i.\,e. at a Hartmann number larger than the
regular flow at $\Ha=0.7$ presented in this section, whereas the
bottom plot corresponds to the chaotic flow at $\Ha=0.63$ with larger
variation of $f$ in Fig.~\ref{fig:Lask_mov_RT}.  Although for the
chaotic flow the moving spectra exhibit some irregular bands around
secondary frequencies, the frequency of largest amplitude remains
constant to $f=8.9$ at $\Ha=0.63$ (also $\Ha=0.67$), which is the same
value obtained for the regular flow, at significantly different
$\Ha=0.74$. For the figure a time window of $T=10$ is used, however,
for the other values of $T=5,20$, and $40$, the results remain
basically unchanged. This confirms the results presented in the
previous section, i.\,e., for chaotic flows with $\Ha$ near the onset
of chaos a highly accurate determination of the frequency of maximum
amplitude is required to detect the chaotic nature of these flows
because the time fluctuations of this frequency can be very small.

\section{Summary and conclusions}
\label{sec:sum}

\begin{table*}[t!]
  \caption{Mean frequency $\overline{f}$ with maximum amplitude, and
    its absolute difference $\varepsilon_f=f_{\max}-f_{\min}$. The
    time dependent frequency of maximum amplitude $f$ is computed on a
    time window $T$ from the time series of the radial velocity picked
    up at the point $(r,\theta,\varphi)=(r_i+0.5d,\pi/8,0)$. This
    frequency is associated to the drifting behaviour of the
    waves. The superscript $\ast$ indicates a regular solution,
    otherwise the solution is chaotic.}
  \label{table:fmean_vr}
  \hspace{5.5cm}Newhouse-Ruelle-Takens    \hspace{2.5cm}Feigenbaum  \\[-0.6cm]
\center  
\scalebox{1.}{
\begin{tabular}{|l|l|lll|lll|}    
\hline
$T$   & $\Ha$          & $0.7^*$          & $0.67$           & $0.63$            & $3.425^*$        & $3.4$            & $0.7$\\ 
\hline
$1$   & $\overline{f}$ & $42.13$          & $42.13$          & $42.13$           & $42.01$          & $42.01$          & $40$ \\
$1$   & $\varepsilon_f$   & $3\times 10^{-2}$ & $4\times 10^{-2}$ & $6\times 10^{-2}$ & $4\times 10^{-2}$ & $5\times 10^{-2}$ & $20$ \\
\hline
$2.5$ & $\overline{f}$ & $42.131$         & $42.13$          & $42.13$           & $42.014$         & $42.013$         & $42$ \\
$2.5$ & $\varepsilon_f$   & $4\times 10^{-3}$ & $~~10^{-2}$       & $2\times 10^{-2}$ & $4\times 10^{-3}$ & $6\times 10^{-3}$ & $17$ \\
\hline
$5$   & $\overline{f}$ & $42.1303$        & $42.132$         & $42.134$          & $42.0144$        & $42.0131$        & $42.29$ \\
$5$   & $\varepsilon_f$   & $2\times 10^{-4}$ & $4\times 10^{-3}$ & $9\times 10^{-3}$ & $2\times 10^{-4}$ & $5\times 10^{-4}$ & $8\times 10^{-2}$ \\
\hline
$10$  & $\overline{f}$ & $42.1304$        & $42.132$         & $42.134$          & $42.01448$       & $42.01337$       & $42.29$ \\
$10$  & $\varepsilon_f$   & $~~10^{-4}$       & $2\times 10^{-3}$ & $5\times 10^{-3}$ & $2\times 10^{-5}$ & $6\times 10^{-5}$ & $6\times 10^{-2}$ \\
\hline
$20$  & $\overline{f}$ & $42.13057$       & $42.1316$        & $42.134$          & $42.014481$      & $42.01337$       & $42.29$ \\
$20$  & $\varepsilon_f$   & $9\times 10^{-5}$ & $6\times 10^{-4}$ & $4\times 10^{-3}$ & $~~~~~~0$        & $2\times 10^{-5}$ & $3\times 10^{-2}$ \\
\hline
$40$  & $\overline{f}$ & $42.13082$       & $42.1313$        & $42.134$          & $42.014481$      & $42.01336$       & $42.295$ \\
$40$  & $\varepsilon_f$   & $3\times 10^{-5}$ & $3\times 10^{-4}$ & $2\times 10^{-3}$ & $~~~~~~0$        & $~~~10^{-5}$      & $2\times 10^{-2}$ \\
\hline
\end{tabular}}
\end{table*}

\begin{table*}[t!]
  \caption{Mean frequency $\overline{f}$ with maximum amplitude, and
    its absolute difference $\varepsilon_f=f_{\max}-f_{\min}$. The
    time dependent frequency of maximum amplitude $f$ is computed on a
    time window $T$ from the time series of the volume averaged
    kinetic energy of the $m=2$ component of the flow. This frequency
    is associated to the modulation behaviour of the waves. The
      superscript $\ast$ indicates a regular solution, otherwise the
      solution is chaotic.}
  \label{table:fmean_mener}
  \hspace{6.cm}Newhouse-Ruelle-Takens    \hspace{2.cm}Feigenbaum  \\[-0.6cm]
  \center
\scalebox{1.}{
\begin{tabular}{|l|l|lll|lll|}    
\hline
$T$   & $\Ha$          & $0.7^*$          & $0.67$           & $0.63$           & $3.425^*$       & $3.4$            & $0.7$\\ 
\hline
$1$   & $\overline{f}$ & $8.92$           & $8.9$            & $8.9$            & $4.5$           & $4.5$            & $4$ \\
$1$   & $\varepsilon_f$   & $7\times 10^{-2}$ & $2\times 10^{-1}$ & $2\times 10^{-1}$ & $~~10^{0}$      & $10^{0}$          & $25$ \\
\hline
$2.5$ & $\overline{f}$ & $8.92$           & $8.93$           & $8.9$            & $4.59$           & $4.59$           & $25$ \\
$2.5$ & $\varepsilon_f$   & $3\times 10^{-2}$ & $8\times 10^{-2}$ & $~~10^{-1}$      & $5\times 10^{-2}$ & $8\times 10^{-2}$ & $3$ \\
\hline
$5$   & $\overline{f}$ & $8.916$          & $8.93$           & $8.94$           & $4.589$           & $4.59$          & $25$ \\
$5$   & $\varepsilon_f$   & $~~10^{-3}$       & $3\times 10^{-2}$ & $9\times 10^{-2}$ & $2\times 10^{-3}$ & $~~10^{-2}$      & $3$ \\
\hline
$10$  & $\overline{f}$ & $8.91589$        & $8.93$           & $8.93$           & $4.58856$         & $4.589$          & $6$ \\
$10$  & $\varepsilon_f$   & $9\times 10^{-5}$ & $2\times 10^{-2}$ & $4\times 10^{-2}$ & $9\times 10^{-5}$ & $4\times 10^{-3}$ & $3$ \\
\hline
$20$  & $\overline{f}$ & $8.915878$       & $8.93$            & $8.93$           & $4.58852$        & $4.589$          & $25$ \\
$20$  & $\varepsilon_f$   & $2\times 10^{-6}$ & $~~10^{-2}$       & $2\times 10^{-2}$ & $2\times 10^{-5}$ & $~~10^{-3}$       & $3$ \\
\hline
$40$  & $\overline{f}$ & $8.9158783$      & $8.926$          & $8.93$            & $4.588518$        & $4.5893$         & $25$ \\
$40$  & $\varepsilon_f$   & $~~~~~0$         & $6\times 10^{-3}$ & $~~10^{-2}$        & $2\times 10^{-6}$ & $6\times 10^{-4}$ & $4$ \\
\hline
\end{tabular}}
\end{table*}

\begin{figure}[h!]
\begin{center}
\hspace{0.cm}  \includegraphics[scale=1.5]{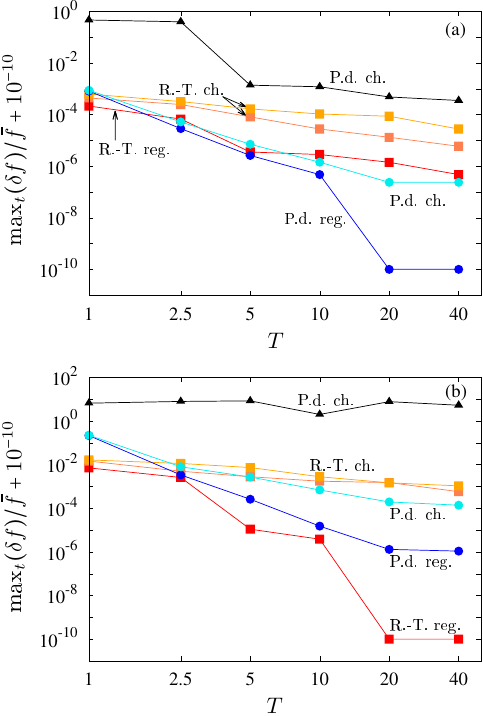}
\end{center}  
\caption{Relative difference $\text{max}(\delta
  f(t))/~\overline{\hspace{-1.mm}f},~0\le t\le 100,$ with $\delta
  f(t)=|f(t+T)-f(t)|$, versus the time window $T$ for the
  Newhouse-Ruelle-Takens (R.-T., full squares), Feigenbaum (P.d., full
  circles and triangle) chaos (ch.) scenarios. Regular (reg.)  flows
  for the R.-T. and P.d. scenarios are considered as well. The time
  series correspond to (a) the volume averaged kinetic energy of the
  $m=2$ component of the flow and (b) the radial velocity picked up at
  the point $(r,\theta,\varphi)=(r_i+0.5d,\pi/8,0)$.}
\label{fig:T_derfreq}   
\end{figure}

The present study is based on very long high order time integrations
of the MSC equations, with a discretized system of $O(10^5)$ degrees
of freedom. Specifically the DNS, on a spherical shell with an aspect
ratio $\chi=0.5$, cover $100$ viscous time units at a Reynolds number
$\Ree=10^3$. This represents around $1.6\times 10^4$ inner sphere
rotation periods, which is a value two orders of magnitude larger than
the achieved by previous studies in the field
(e.\,g. \citet{Hol09,Kap14}).

The time series of local (the radial velocity at a point inside the
shell) and global (a volume-averaged kinetic energy) measures have
been analyzed using Laskar's algorithm for the determination of
fundamental frequencies (\citet{Las93}). The accuracy of the method
(down to $10^{-7}$ in relative values) is estimated using a periodic
flow (a rotating wave) from which the frequency can be obtained using
a Newton-Krylov procedure (\citet{GaSt18}).

Several regular and chaotic flows are selected for the analysis. At
first one regular and two chaotic flows from the Feigenbaum scenario
(\citet{Fei78}) are investigated. Similarly, one regular and two
chaotic flows representing the Newhouse-Ruelle-Takens scenario
(\citet{NRT78}) are selected as well. These two routes to chaos were
confirmed in \citet{GSGS20} by computing the corresponding solution
branches and performing a Poincar\'e section analysis. In this paper
we extend the previous study of \citet{GSGS20} by investigating time
dependent frequency spectra. Following the work of \citet{Las93b}, the
frequency of maximum amplitude $f$ is computed on several time windows
to study the time dependence of $f(t)$ and $\delta
f(t)=|F(t+T)-f(t)|$. This helps to confirm the existence of chaos and
to estimate the diffusion of the orbit in the phase space.

The results are summarized in Tables~\ref{table:fmean_vr}
and~\ref{table:fmean_mener} and in figure~\ref{fig:T_derfreq}. From
the tables as well as from the figure it can be concluded that a
minimum time window of $T=5$ viscous time units should be used to
obtain reliable results that allow an identification of chaos. This is
because for small time windows, $T\le 5$, the fluctuations of $f(t)$
(either measured by $\varepsilon_f$, or by $\text{max}(\delta
f(t))/~\overline{\hspace{-1.mm}f},~0\le t\le 100,$) observed for the
regular solutions are of the same order of magnitude than those
observed for the chaotic flows.

The range of variation of the frequency, $\varepsilon_f =
f_{\max}-f_{\min}$, for the local measure (radial velocity;
table~\ref{table:fmean_vr}) is significantly smaller than that for the
global one (volume-averaged kinetic energy;
table~\ref{table:fmean_mener}). This is also true when considering the
relative difference $\text{max}(\delta
f(t))/~\overline{\hspace{-1.mm}f},~0\le t\le 100,$ displayed in
Fig.~\ref{fig:T_derfreq}. The latter evidences different diffusion
rates in the phase space as the measured frequencies are obtained from
different components (total or volume-averaged) of the flow.

The classical Fourier transform is not accurate enough to detect small
frequency changes over time $|\delta f|< 2/T=10^{-2}$ for the
considered observation windows. This motivates the use of an
optimization method like Laskar's, which finds the most probable
frequency due to a Newton method and (successive) single mode
elimination. The result is a more precise estimate of the dominant
frequency; we have shown that for the Feigenbaum and
Newhouse-Ruelle-Takens scenarios this is key for identifying the onset
of those chaotic flows, which exhibit small variations of $f$.

We stress that a rigorous confirmation of a strange attractor would
require the computation of the leading Lyapunov characteristic
exponent (LCE) (e.\,g. \citet{EcRu85}). This is beyond of the scope of
this study as it would require long time runs of the MSC system
coupled with its first variationals (see \citet{BGGS80} for details),
which is a challenging computational task. We note that the validity
for applying time dependent frequency analysis in chaotic systems has
been already tested by \citet{LFC92} and \citet{GMS10} against the
computation of LCE. The latter studies already noted that frequency
analysis seems to require a shorter time evolution than the
computation of LCE to detect a strange attractor, which is a key issue
for large dimensional systems.

As commented in the introductory section there exist tools which are
able to estimate LCE from a time series (see \citet{HKS99} and
references therein) but these tools require to tune several input
parameters (e.\,g. \citet{AKEDBK18b}) and thus are more sophisticated
than Laskar's method, which only depends on the precision of the
frequency computation. This is also true for other time series methods
such as those based on wavelet transforms (\citet{Dau91,Mey93}) which
decompose the signal into a set of orthogonal basis functions,
localized in the time-frequency domain. Wavelet methods have been
successfully used for the analysis of chaotic solutions by
\citet{StWo99} (also by \citet{SCWI13} for a magnetized plasma
experiment) but they require to select the wavelet type, filter and
length, which are key parameters to be tuned to obtain an accurate
analysis (see \citet{Zha_etal16} in the case of neuron activity
signals).

A remarkable result is that for all types of flows, the frequency
corresponding to the mean azimuthal drift (inferred from the radial
velocity) remains nearly constant and only oscillates less than
$0.2\%$ with respect to its mean value giving rise to very small
($<10^{-2}$) diffusion rates. This is especially surprising in the
case of the highly oscillatory chaotic flow from the Feigenbaum
scenario at $\Ha=0.7$, as the frequency corresponding to the main time
scale of a volume-averaged quantity can vary more than one order of
magnitude. We have tested other chaotic flows in the same branch as
well as all other classes of chaotic flows found by \citet{GSGS20} and
the results are similar. The main conclusion is that the azimuthal
drift behaviour of flows at moderate Reynolds number $\Ree=10^3$ is
strongly robust, even for highly oscillatory chaotic flows.

As found in \citet{GSGS20}, for each class of flows with azimuthal mode
$m_{\max}$, mostly contributing to the kinetic energy, the frequency
associated with the azimuthal drift was very close to that of the
unstable rotating wave with azimuthal symmetry $m_{\max}$, at the same
Hartmann number. With the present analysis we have demonstrated that
this frequency is indeed quite robust even when considering long time
integrations. Thus, unstable rotating waves provide a good description
of the main time scale of the MSC flow at moderate $\Ree=10^3$ and
$\Ha<6$.

The present study sheds light on the analysis of future HEDGEHOG
experiments at $\Ree=10^3$ and $\Ha<4$, which corresponds to the
radial jet instability regime. The experiment is designed to
effectively work for low $\Ha$ (see \citet{KKSS17}) assuming an error
of about $1$\% in the selection of the parameters (see also
\citet{OGGSS20}). The analysis of the DNS points to the difficulty of
distinguishing regular and chaotic flows in the experiment using time
dependent spectral analysis since highly accurate computation of the
main frequency (provided by Laskar's algorithm) is required.

The key issue is that chaotic DNS exhibit small time fluctuations of
the main frequency of the flow velocity, which require large enough
observation windows of $T\ge 5$, to be distinguishable from the
intrinsic numerical fluctuations associated to the approximation of
the frequency, which are also present in the case of regular
solutions. With a kinematic viscosity of the eutectic alloy GaInSn of
$\nu=3.4\times 10^{-3}$\nolinebreak cm$^2$s$^{-1}$ (\citet{PSEGN14})
the time scale in seconds is $t^*=t d^2/\nu=5.96\times 10^{3} t$, $t$
being the dimensionless time and the gap width $d=r_o-r_i= 9\mbox{ cm}
- 4.5\mbox{ cm} = 4.5 \mbox{ cm}$.  Thus $T\ge 5$ represents around 8
hours of the HEDGEHOG experiment, which is almost the limit of a
typical experimental run (up to 10 hours) because of the degradation
of the signal quality (\citet{OGGSS20}). This means that observational
time windows of maximum size $T\sim 2$ can be considered in the
experiment which are impractical for detecting chaotic flows if the
analysis of the main frequency obtained from a velocity measurement is
performed.

By analyzing the volume-averaged kinetic energy, and not directly the
flow velocity, we have shown that the main frequency can vary several
orders of magnitude (even for small time windows of $T=1$) in the case
of a chaotic flow belonging to the Feigenbaum scenario at $\Ha=0.7$
(see Table~\ref{table:fmean_mener}). These chaotic flows could be
detected in the experiment provided that the time-dependent spectral
analysis is performed to a secondary frequency of the flow reflecting
the modulation of volume-averaged properties. This can be done within
the HEDGEHOG measurement setup as the frequency spectrum can be
computed independently for the $m=0,1,2,3,4$ and $5$ azimuthal wave
number components of the flow (see \citet{OGGSS20} for details). The
analysis of the volume-averaged kinetic energy does not work for
detecting the onset of chaotic flows if time windows of $T<5$ are
used, neither for the Feigenbaum nor the Newhouse-Ruelle-Takens
scenarios.

\section*{Acknowledgments}
F. Garcia kindly acknowledges the Alexander von Humboldt Foundation
for its financial support.  This project has also received funding
from the European Research Council (ERC) under the European Union’s
Horizon 2020 research and innovation programme (grant agreement No
787544).





\end{document}